\documentclass[12pt,preprint]{aastex}
\usepackage{graphicx,color}
\def\red#1 {\textcolor{red}{#1}}
\newcommand\etal{{\it et al.}}
\begin{document}

\title{A Deep 0.3-10 keV Spectrum of the H$_{2}$O Maser Galaxy IC 2560}
\author{Avanti Tilak and Lincoln J. Greenhill}\affil{Harvard-Smithsonian Center for Astrophysics, 60 Garden Street, Cambridge, MA 02138}\email{atilak@cfa.harvard.edu}
\author{Chris Done}\affil{Department of Physics, Durham University, Durham, UK}
\and
\author{Grzegorz Madejski\altaffilmark{1}}\affil{Stanford Linear Accelerator Center, Menlo Park, CA 94025}
\altaffiltext{1}{Kavli Institute for Particle Astrophysics and Cosmology, Stanford, CA 94305}

\begin{abstract}
We present a new XMM-Newton spectrum of the Seyfert 2 nucleus of IC 2560, which hosts H$_{2}$O maser emission from an inclined Keplerian accretion disk. The X-ray spectrum shows soft excess due to multi-temperature ionized plasma, a hard continuum and strong emission features, from Mg, Si, S, Ca, Fe and Ni, mainly due to fluorescence. It is consistent with reflection of the continuum from a mostly neutral medium and obscuration due to a high column density, $>$ 10$^{24}$ cm$^{-2}$. The amplitude of the reflected component may exceed 10\% of the central unobscured luminosity. This is higher than the reflected fraction, of a few percent, observed in other Seyfert 2 sources like NGC 4945. We observe an emission line at 6.7 keV, possibly due to FeXXV, undetected in previous Chandra observations. The absorption column density associated with this line is less than 10$^{23}$ cm$^{-2}$, lower than the obscuration of the central source. We hypothesize that this highly ionized Fe line emission originates in warm gas, also responsible for a scattered component of continuum emission that may dominate the spectrum between 1 and 3 keV. We compare X-ray and maser emission characteristics of IC 2560 and other AGN that exhibit water maser emission originating in disk structures around central engines. The temperature for the region of the disk associated with maser action is consistent with the expected 400-1000K range. The clumpiness of disk structures (inferred from the maser distribution) may depend on the unobscured luminosities of the central engines.

\end{abstract}
\keywords{Galaxies: Active, Galaxies: Seyfert, Nuclei, Individual:-- (IC 2560), Galaxies: Accretion Disks, Masers}
\section{INTRODUCTION}

IC2560 is a nearby spiral ($cz\sim 2925$\,km\,s$^{-1}$) that is classified as a Seyfert 2 \citep{fai86} and known to host 22\,GHz H$_2$O maser emission originating in an accretion disk around the central engine \citep{ishi01}.  Here we present analysis of a deep (archival) XMM-Newton EPIC observation of this active galactic nucleus (AGN).   Previous observations of IC 2560 in the X-ray band using ASCA GIS \citep{ishi01} and Chandra ACIS-S \citep{iwa02,mad06} both pointed to a reflection dominated spectrum and high obscuring column density.

IC 2560 is a relatively weak X-ray source, with an observed 2-10 keV flux of 3.3$\times$10$^{-13}$ erg s$^{-1}$ cm$^{-2}$ that originates predominantly from a central source \citep{mad06}. Interpretation of early GIS and ACIS-S observations was constrained by relatively limited sensitivity and energy resolution, in particular above 6 keV, where the bulk of the emission due to nuclear activity escapes into our line of sight. The EPIC instrument provides $\sim$5$\times$ larger collecting area at 6 keV and $\sim$20\% finer energy resolution (160 eV) than ACIS-S (chip 3). Although dominated by the central point source, flux detected with EPIC (1.1 kpc pixel$^{-1}$ at 1 keV) will include contribution from extended and off-nuclear emission components. As such, we note that the nuclear structure in IC 2560 appears to be relatively complex, as may be anticipated for a type-2 object. Hubble Space Telescope ACS/HRC observations at 3300 \AA\ detect patchy emission with multiple peaks within the central $\sim$ 200 pc \citep{mun07}, presumably due to dust. Relatedly, population synthesis models fit to optical spectra, obtained with the ESO 1.5m (3470-5450 \AA) for the central $\sim$ 300 pc, are suggestive of active star formation during the last 1.5-10 Myr \citep{fer04}. Ground-based spectroscopic study has also detected asymmetry in the [OIII] line profile (a blue excess) that may be indicative of outflow at $\sim$ 100 km s$^{-1}$ projected on the sky plane in a 300$\times$400 pc central patch \citep{schu03}. Because the inclination of the galaxy is 63$^\circ$, and the inclination of the central engine (as inferred from  maser emission) may be even closer to edge-on, the actual flow speed may be larger.

The observed H$_2$O maser spectrum exhibits emission close to the systemic velocity as well as blue and red-shifted complexes, offset, approximately symmetrically, by $\pm$ 200 - 420 km s$^{-1}$ \citep{bra03}. This pattern is an indicator of emission from a rotating edge-on disk structure (e.g., NGC 4258; \cite{nak93}). \cite{ishi01} reported line-of-sight accelerations for the ``systemic emission'' and mapped its emission distribution using Very Long Baseline Interferometry (VLBI). Based on the assumption that the emission traced a thin, rotating, circular disk that is observed edge-on, they used the velocities of high-velocity emission to estimate disk radius and dynamical mass. Follow-up VLBI observation by Greenhill \etal (2008, in preparation) enabled mapping of both systemic and high-velocity emission, thus tracing the disk structure, demonstrating Keplerian rotation of material around the central engine in a relatively thin distribution, and enabling more certain modeling of the accretion disk geometry and estimation of the central mass. The observed molecular gas lies at radii 0.08 - 0.27 pc and the inferred enclosed mass is 2.9$\times$10$^6$M$_\odot$ with $\sim$ 20\% accuracy. The thinness of disk (h/r $\ll$1) as well as presence of H$_2$O emission suggest disk material that is relatively cool ($\la$10$^3$ K).

Due to generally high extinction toward Seyfert-2 central engines, X-ray and radio observations together are especially useful in the study of the physical processes that occur at radii $\la$1 pc. Measurement of X-ray spectra enable estimation of intrinsic luminosity, obscuring column density, temperature and abundance. Time monitoring facilitates estimation of size scales for high-energy phenomenon. VLBI observations of H$_2$O maser emission enable mapping of accretion disk geometry and orientation to the line of sight, and estimation of orbital radii for molecular material, central engine mass and Eddington luminosity (\cite{gre07} and references therein). The presence of maser emission implies the existence of a reservoir of relatively cool material and at the same time, a heating mechanism to maintain population inversion. The necessary energy may be  imparted by viscous heating within the disk (e.g., \citep{des98}) or external irradiation \citep{neu94,wat94}. However, in either case, survival of molecular gas demands a shielding column from UV and X-ray emission generated by, e.g., the central engine and coronae. This column may arise in outflowing material, disk surface layers, or disk material at small radii, thus providing additional quantitative constraints on overall structure around the central engine and energetics. In section 2, we discuss the data reduction of the EPIC data, while section 3 focuses on analysis of the source spectrum. Estimation of the intrinsic emission of the central engine (i.e., corrected for absorption along line of sight) is discussed in section 4. X-ray properties of IC 2560 and other extragalactic H$_2$O maser sources believed to originate in accretion disks are considered in section 5; there we put IC 2560 into context and we discuss possible trends that may indicate connection between characteristics estimated from VLBI observations of masers and X-ray observations of their host AGN. Throughout the paper we adopt H$_{0}$=70 km s$^{-1}$ Mpc$^{-1}$.
\section{DATA REDUCTION AND ANALYSIS}

IC 2560 was observed using the XMM-Newton EPIC instrument on 2003 December 2. Data reduction was carried out using SAS version 7.0.0. PN and MOS datasets were processed through the standard pipeline, using updated CCF files and SAS tasks $\bf{epchain}$ and $\bf{emchain}$, respectively. Light curves extracted from the data do not indicate any of the cameras observed flares and hence no time intervals required filtering. The final exposure time was 72.2 ks for the PN camera, 80.5 ks for MOS1 and 80.6 ks for MOS2.

We extracted the source spectrum from a circular patch, 15$\arcsec$ (2.95 kpc) in radius, centered on the source coordinates in the binned image (binsize 80). Tests using the SAS task $\bf{epatplot}$ for PN data show good agreement between the expected and observed distribution of counts for the single, double, triple and quadruple event patterns. The source spectrum is thus unaffected by pileup. We used a nearby source-free circular region of radius 40$\arcsec$ to extract the background spectrum, selected so that the source and background are equidistant from the readout node. After normalizing for area, the PN spectrum had a 3.1\% contribution from the background, while the MOS1 and MOS2 spectra show a lower contribution from the background, approximately 2\%. The source and background spectra were binned so that each bin contained at least 20 counts. Models were fitted to the spectrum between energies 0.3-13 keV. 

\begin{figure}
\rotate
\begin{center}
\scalebox{0.75}[0.75]{\plotone{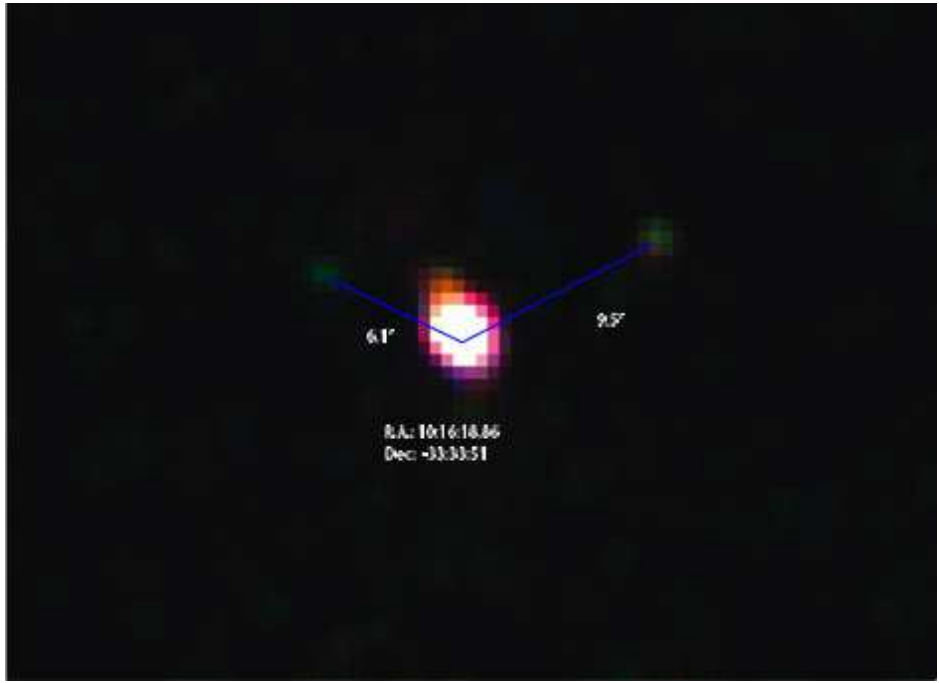}}
\caption{Chandra image of IC 2560. North is up. Energies 0.3-1.5keV represented by red, energies 1.5-2.5 keV by green and 2.5-8 keV energy represented by blue colors.Diffuse emission from galactic plane of IC 2560 is visible as are two discrete sources of emission, approximately 9.5$\arcsec$ from the central source, about 50$^o$ from the north and 6.1$\arcsec$ from the center, about -20$^o$ from the north. The image has been smoothed using a Gaussian function with radius 2 pixels, to suppress statistical noise. Coordinates indicate brightest pixel in image\label{chand}} 
\end{center}
\end{figure}

The host galaxy has a semi-major axis of approximately 90$\arcsec$ \citep{dev91}. We used archival Chandra data to quantify the contamination of the XMM spectrum, due to large-scale emission from the host galaxy. The Chandra image (from the 2005 observation discussed by \citet{mad06}, Figure \ref{chand}) shows X-ray emission from two discrete sources, approximately 9.5$\arcsec$ and 6.1$\arcsec$ from the brightest pixel (within 0.8$\arcsec$ of the optical center, the offset is comparable to the positional accuracy of Chandra at 99\% limit and slightly higher than the 90\% limit of 0.6$\arcsec$). The discrete emission appears to be dominated by soft and medium energy X-rays ($\le$ 2.5 keV). Diffuse emission in the plane of the galaxy is also visible. A circular region of radius 15$\arcsec$ (size of the extraction region for the PN and MOS spectra), centered on the source yields 2201 counts, and 601 counts excluding the central 4$\arcsec$ for energies between 0.3 keV and 10 keV. The Chandra background as measured from a 15$\arcsec$ circle, placed approximately 90$\arcsec$ from the center (comparable to the host galaxy semi-major axis) contains 435 counts. Thus, after background subtraction, we find the large scale emission from the host galaxy contributes approximately 9\% to the source spectrum in the XMM observation.
\section{SPECTRAL FITTING}

\begin{figure}
\begin{center}
\scalebox{0.65}[0.65]{\includegraphics[angle=-90]{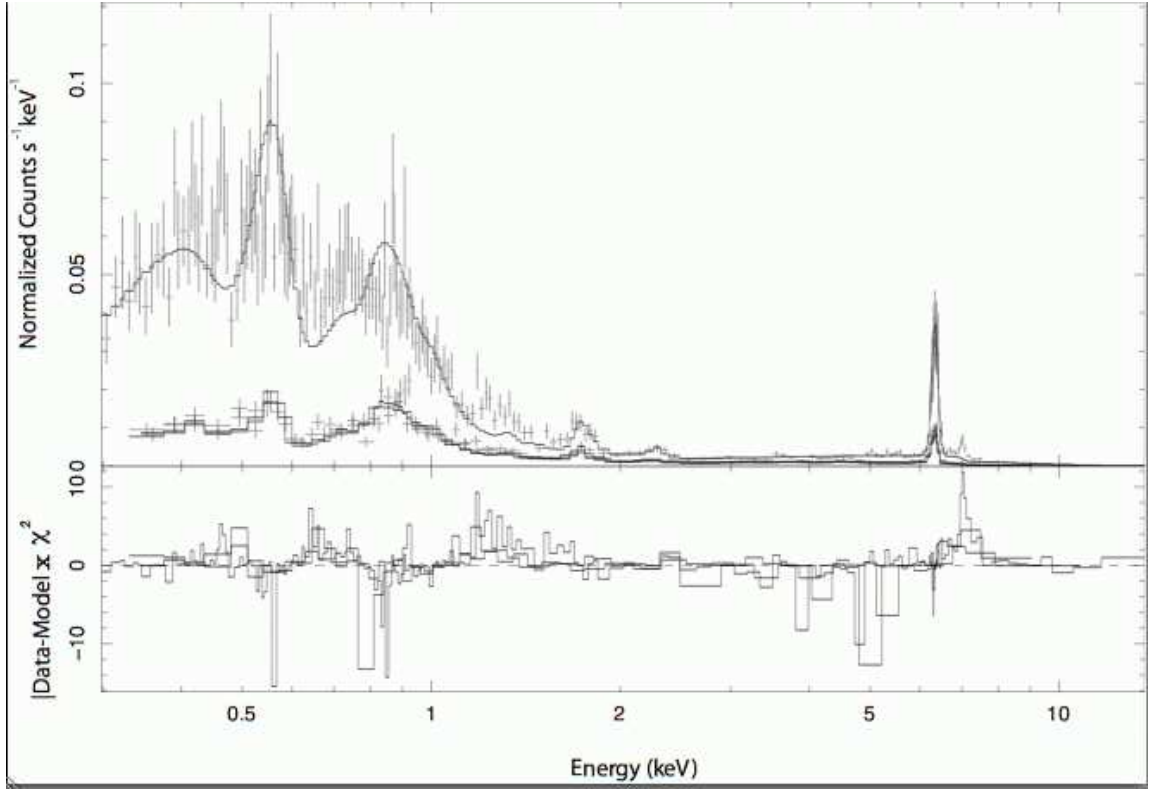}}
\caption{Composite PN and MOS data, binned with 20 cts/bin, fit to the Chandra best-fit model (top) and $\Delta$$\chi$$^{2}$ residual (bottom). The model is described in Table 1 as Chandra-1.\label{chandra}}
\end{center}
\end{figure}

All spectral fitting  was carried out using XSPEC version 11.3 \citep{arn96}. The spectrum shows all the key features observed in the Chandra spectrum \citep{mad06} i.e. a hard continuum above 3 keV with a strong Fe K$\alpha$ fluorescence line at 6.4 keV, both indicative of a reflection origin in material far from the black hole. Below 2 keV there is additional soft emission, with multiple lines superimposed, that may be due to a complex mix of photoionized and/or collisionally ionized plasma. We use the best-fit model of \citet{mad06} as a first approximation, consisting of  two MeKaL collisionally ionized plasmas to describe the soft X-ray emission, a reflected component of the (hidden) intrinsic power-law emission modeled by PEXRAV \citep{mag95} at higher energies, with a fixed inclination angle of 60$^\circ$, together with the associated neutral fluorescence lines from Fe, Si and S K$\alpha$ with rest frame energies of 6.4, 1.75 keV and 2.3 keV respectively \citep{bea67}. We refer to this as Chandra-1, and use this as a starting point to fit the PN+MOS EPIC data from XMM-Newton (Table \ref{tab-spec}). Fitting this to the data gives a good description to the overall spectral shape, but the better signal-to-noise XMM-Newton data show significant line-like residuals (Figure \ref{chandra}). The strongest of these are at $\sim$~7 and 1.2~keV, corresponding to neutral Fe K$\beta$ and Mg K$\alpha$, respectively, but there are also significant detections of neutral K$\alpha$ emission from Ni and (marginally) Ca at 7.4 and 3.6~keV, as well as {\em ionized} K$\alpha$ emission from He-like Fe at 6.7 keV. 

\begin{figure}
\begin{center}
\scalebox{1}[1]{\plotone{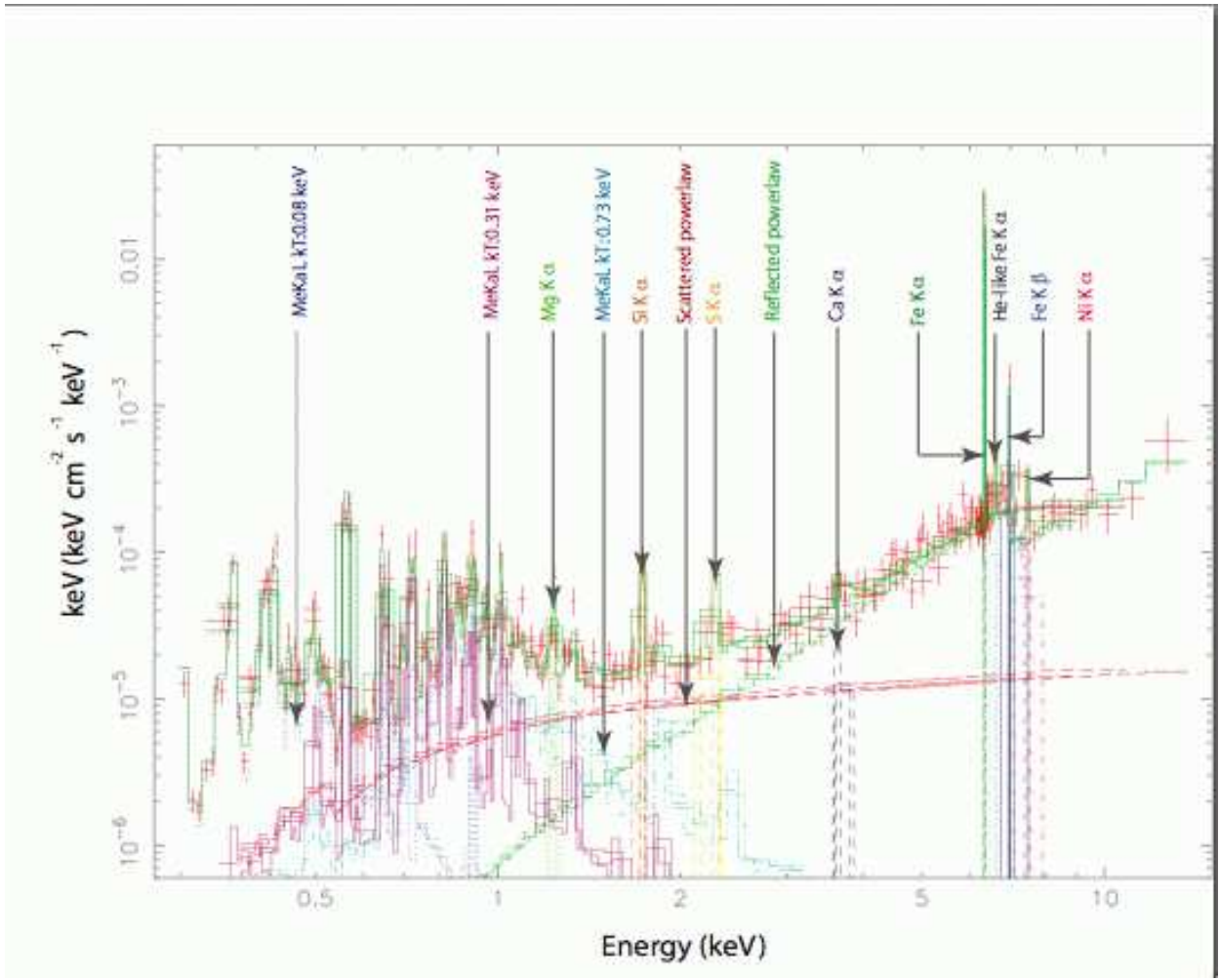}}
\caption{Best-fit unfolded model and spectrum for XMM PN and MOS data, binned with 20 cts/bin, summarized in Table 1, XMM-bf. Both PN and MOS unfolded data are shown by red crosses. \label{xmm-eeuf}}
\end{center}
\end{figure}

We also tested for the presence of other lines. Neutral K$\alpha$ fluorescence lines from Ar (2.95 keV),  O (0.55 keV) and Ne (0.84 keV) could also be present from the reflector. These are not significantly detected, but the upper limits to the equivalent widths are consistent with expectations from solar abundances \cite{matt97}. Similarly the data only give upper limits to the emission from other ionized lines which might be expected to accompany the He-like Fe at 6.7~keV such as H-like Fe K$\alpha$ (6.955), He-like Ni K$\alpha$(7.8 keV) blended with He-like Fe K$\beta$(7.88 keV) and H-like Ni K$\alpha$(8.09 keV). However, there is clear evidence for further ionized line emission at low energies, which is well matched by including a third plasma component. While this model (XMM-1, 3 MeKaL, 8 Gaussian lines and a PEXRAV continuum, Table \ref{tab-spec}) matches the data fairly well ($\chi^2$/d.o.f=358.7/312), it is physically inconsistent. The He--like Fe line at 6.7 keV is either from a photoionized plasma, in which case it should be accompanied by a scattered power-law continuum, or from collisionally ionized plasma, in which case it should be produced by an additional MeKaL component. However, the required temperature of this 4th hot plasma emission is 4.4$^{+2.9}_{-1.8}$ keV (XMM-2, $\chi^2$/d.o.f.= 322.0/311, Table \ref{tab-spec}), which seems high for starburst emission. It could indicate a violent interaction between the interstellar medium and an outflow \citep{vei97}, however, the detected outflow velocity \citep{schu03} of 100 km s$^{-1}$ is fairly low. Instead we favor the scattered power-law for a photoionized origin of the 6.7 keV line, and note the resulting (marginally) better $\chi^2$/d.o.f = 317.4/311. This gives our best-fit model XMM-bf (Figure \ref{xmm-eeuf}), to the XMM PN+MOS data, as described in Table 1. 

\subsection{Line Emission}

The lines that contribute significantly to the spectrum and their equivalent widths are summarized in Table \ref{tab-eqw}. The equivalent widths for mostly neutral species are computed with respect to the reflected power-law component. The spectrum has not evolved substantially since the Chandra observations reported by \citet{iwa02} and \citet{mad06}. Specifically, we do not detect any statistically significant variability in the continuum flux in the 2-10 keV range or in the equivalent widths of the three emission lines detected in the Chandra dataset (Table \ref{tab-var}), though S/N is limited. The best-fit value for the 6.4 Fe K $\alpha $ line in the XMM dataset for model Chandra-1 is  close to the lower limit for the same in the Chandra dataset. We attribute this to blending of 6.4 and 6.7 keV emission lines in the Chandra observation. The Fe K$\alpha$ line in the XMM spectrum is marginally resolved and favors an intrinsic width of $\sim$20 eV and $<$50 eV, similar to the Chandra data, and could originate in material at a few thousand km s$^{-1}$, typical of the BLR/inner torus. The emission features at 1.25 keV, 3.6 keV, 6.7 keV, 7.05 keV and 7.47 keV as well as the scattered continuum component are absent from the Chandra model, but these were not required to obtain a reasonable fit to the earlier spectrum. 

We inspected the spectrum for the presence of a 6.33 keV Compton down-scattered shoulder to the 6.4 keV line. To constrain this, we focus on the data above 5 keV, including only the lines within this energy range, together with the PEXRAV and scattered power law as continuum components with photon index fixed (but not norms) to the best-fit value given in Table 1. We included a model \citep{mad06} to describe the shape of the Compton down-scattered shoulder \citep{ill79} and fit this to the data but obtain only an upper limit of about 10 per cent of the narrow K$\alpha$ 6.4~keV line. This is at the low end of what is expected from reflection, indicating a face on view of the reflector (\citet{mat02} assuming the geometry described by \citet{ghi94}) and is in contradiction with the edge-on view of the disk inferred from maser emission. The face on view is also inconsistent with the expected orientation for a Seyfert 2 source under the unification paradigm. 
 
A similarly low Compton shoulder was observed for another Seyfert 2 source, Mrk 3, by \citet{pou05}. However, they point out that this may be due to an error in the registered energy of split-pixel photons, which can be overestimated by up to 20 eV. Following their procedure and selecting only single-pixel events, we fit a Compton shoulder of 1.9$_{-1.9}^{+12.5}$\% for a simultaneous fit to the PN and MOS data. A slightly higher limit was measured at 8.7$_{-8.9}^{+16.2}$\% for PN data alone, though this is probably due to decreased signal-to-noise. 
 
The fitted Compton shoulder height should be viewed with some caution. Detection of the shoulder is close to the limit of what is possible with available data. The inferred intrinsic width for the Fe K$\alpha$ line is comparable to the shift in split pixel registration discussed above. The line width may also affect the detection of the Compton shoulder, due to limitations in modeling. Hence we adopt an upper limit of 25\% on the Compton shoulder but stress that the data at face value are inconsistent with a reflection (or transmission) origin. On the other hand, recent work by \citet{awa07} shows 10\% upper limit for Mrk 3 from higher resolution Suzaku data, consistent with the XMM data at face value. IC 2560 may have a similarly low Compton shoulder, due to a clumpy torus structure as suggested by \cite{awa07}. However in that scenario, it will be very difficult to produce the observed amplitude of the reflected component. 

\subsection{The Soft X-ray emission}
 
The soft X-ray emission dominant below 2 keV is almost certainly due to hot gas. For modeling purposes, we have used three MeKaL components to obtain a good description of this emission. This gives plasma temperatures of 0.1 keV, 0.32 keV and 0.66 keV. Replacing the three components with a single model where the plasma shows a power-law gradient with index 0.001 (best-fit value) in temperature does not give a satisfactory description of the low energy emission ($\chi^2$/d.o.f = 616.8/315).  However, some or even all of the observed emission below 2 keV could also be due to the photoionized scattering medium. Non-nuclear sources in the host galaxy could also have significant contribution to the observed spectrum below 2 keV. We note that the spectral analysis of Chandra data by \citet{mad06} indicated that a collisionally ionized plasma was required even in presence of photoionized gas, in order attain a realistic description of the spectrum. However, the analysis neglected the scattered power-law continuum that must accompany any such photoionized emission, and the complexities of radiative transfer in the 6.7 keV line which can substantially change the strength of the line feature \citep{bia02}. A pure photoionized origin is suggested by the detection of radiative recombination lines associated with OVII at 0.69 keV in RGS data for this source \citep{gua07}, though the signal-to-noise in these data is very limited. The dominance of photoionization in similar Compton thick sources such as NGC 1068 \citep{kin02} and Circinus \citep{mas06} implies it may be an important mechanism in IC 2560 as well. However, given the high density of line emission and low spectral resolution at these low energies, it is difficult to disentangle collisional and photoionization components and measure their relative contributions to the total emission.

\section{ESTIMATE OF UNABSORBED EMISSION}

The obscuration column density for IC 2560 is very high, $\ge$ 10$^{24}$ cm$^{-2}$, as inferred by \citet{iwa02} and \citet{mad06}. This source has not been detected at higher energies (above 10 keV) by the RXTE Slew Survey \citep{rev04} or by SWIFT/BAT \citep{mar05}. If we extrapolate the best-fit spectrum to 200 keV, we find that the total flux from the continuum components in our model would contribute 7.8$\times 10^{-13}$ erg cm$^{-2}$ s$^{-1}$ in the energy range 3-20 keV relevant to the RXTE Slew survey and 2.8$\times 10^{-12}$ erg cm$^{-2}$ s$^{-1}$ in the energy range 14-195 keV corresponding to SWIFT/BAT. These values are not far below the detection limit of 10$^{-11}$ erg cm$^{-2}$ s$^{-1}$ for both surveys, strongly limiting the amount of direct continuum in these bands from the central engine.  

The [OIII] luminosity (L$_{[OIII]}$) can be used to gauge the (unabsorbed) intrinsic 2-10 keV luminosity (L$^I_{2-10}$) of an AGN \citep{pan06}. \citet{gu06} use stellar population synthesis modeling to estimate and subtract the stellar component from observed spectra, to predict a nuclear [OIII] flux of 1.25$\pm$0.05$\times$10$^{-13}$ erg cm$^{-2}$ s$^{-1}$, after a reddening correction of $<$1\% determined from H$_{\gamma}$ and H$_{\beta}$ line ratios. We refit the data from \citet{pan06}, excluding the points with only upper limits, and scaled the parameter uncertainties by $\sqrt(\chi^2$/d.o.f) to obtain Log$_{10}$ L$_{2-10}^I$  = (1.16$\pm$0.04)(Log$_{10}$ L$_{OIII}$-40.22)+(41.57$\pm$0.05). This gives us an approximate L$_{2-10}^I$ =  6.17$_{-0.80}^{+0.90}$$\times$10$^{41}$erg s$^{-1}$. 

In the following discussion, $\it{intrinsic}$ $\it{emission}$ refers to the flux or luminosity (L$^I_{2-10}$) of the central engine that would be visible if there were no obscuration, $\it{direct}$ $\it{component}$ corresponds to intrinsic emission visible after it passes through the obscuring column suffering a reduction in low energy flux. We can also obtain an order of magnitude estimate for the intrinsic luminosity from the IR luminosity of the source. \citet{iwa02} estimate the L$_{IR}$ from the IRAS to be 3$\times$10$^{43}$ erg s$^{-1}$. Assuming a power-law of photon index 1.8 from the disc at 10eV to 100keV and that all the observed IR emission is due to reprocessing of the X-ray emission, this gives intrinsic luminosity L$_{2-10}$=6$\times$10$^{42}$ erg s$^{-1}$. Stellar heating may have a significant contribution to the observed IR flux. The L$^I_{2-10}$ from IR and OIII measurements are within an order of magnitude which is acceptable in light of uncertainties in both estimates.

\begin{figure}
\begin{center}
\scalebox{1}[1]{\plotone{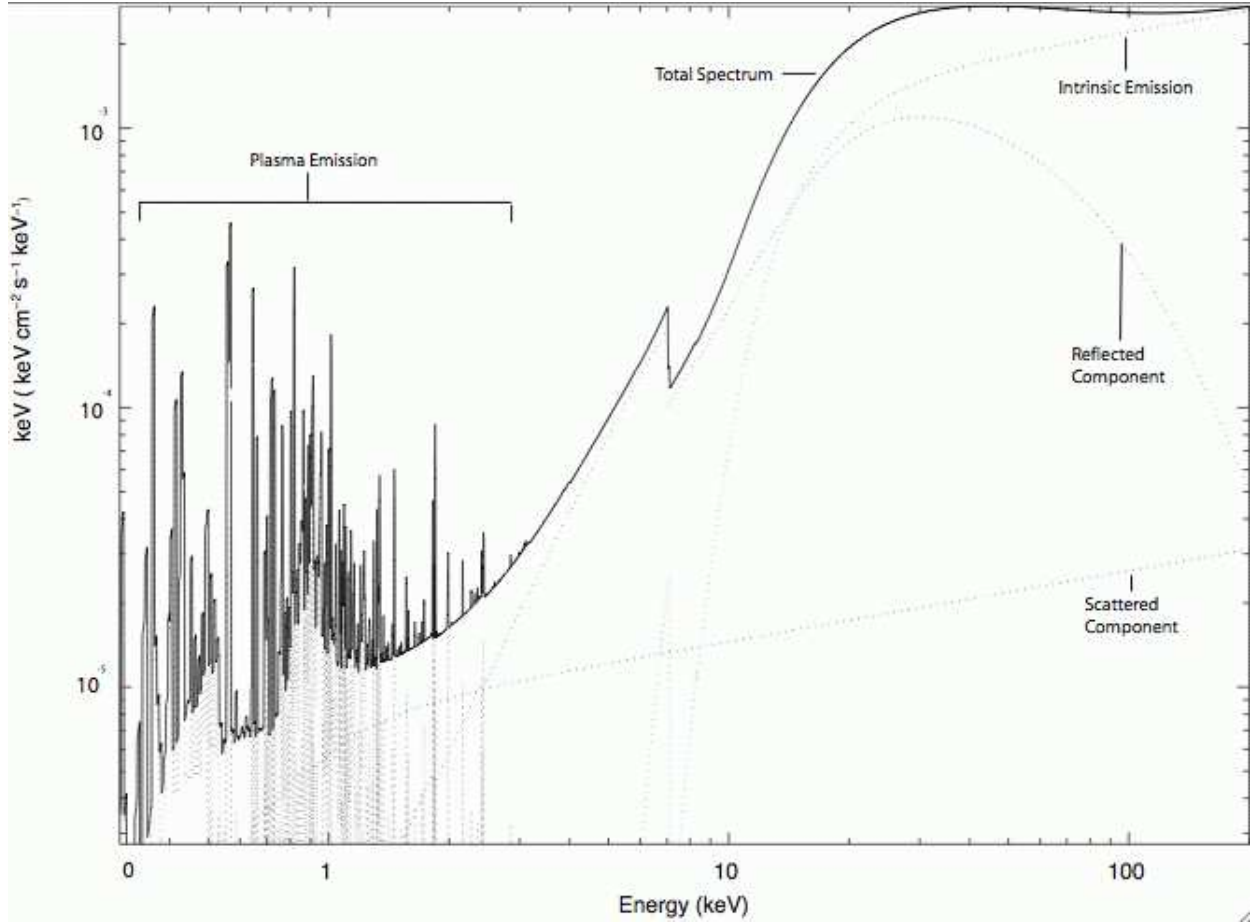}}
\caption{Continuum spectrum extrapolated to 200 keV, with contribution from ionized plasma, scattered and reflected components as well as absorbed intrinsic emission of source. Absorption column density for intrinsic emission set to 3$\times$10$^{24}$ cm$^{-2}$.\label{dummy}}
\end{center}
\end{figure}

\begin{figure}
\begin{center}
\scalebox{1}[1]{\plotone{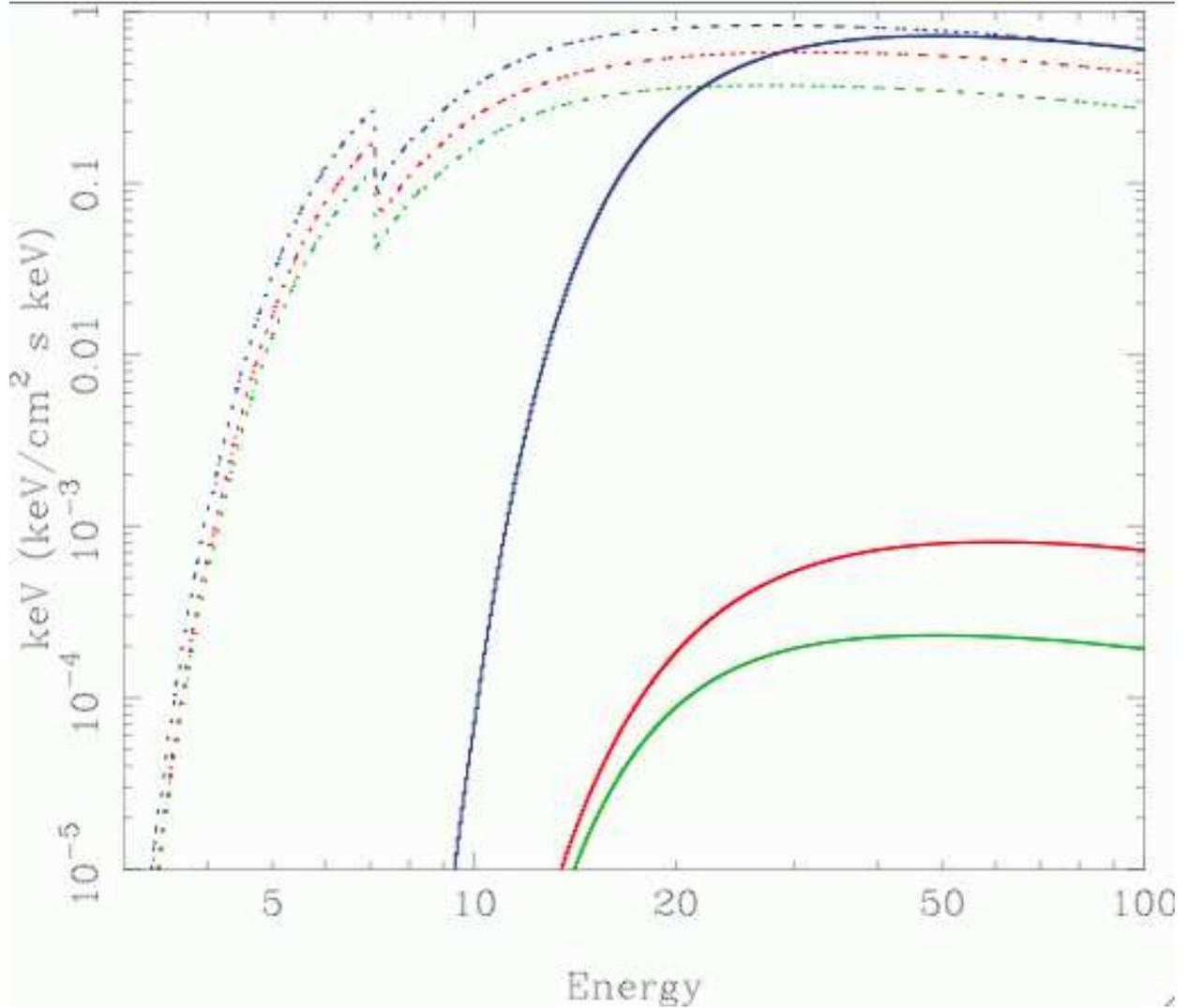}}
\caption{Direct component of intrinsic emission through obscuration columns of density 10$^{24}$ cm$^{-2}$ (dashed lines) and 10$^{25}$ cm$^{-2}$ (solid lines), with only photoelectric absorption (blue), with photoelectric as well as electron scattering out of line-of-sight but not back into line-of-sight, corresponding to a very low covering fraction(green), and combination of both absorption processes including scattering into line-of-sight for unit covering fraction (red).\label{fig-cvd}}
\end{center}
\end{figure}

The estimated L$^{I}_{2-10}$ is consistent with an unabsorbed power-law with photon index 1.75 (best-fit value in XMM-bf) and normalization of 1.3$\times$10$^{-3}$. From this we estimate the reflected solid angle ($\Omega$/2$\pi$) from neutral material $\sim 0.75$ assuming the material is viewed at $60^\circ$ inclination, and the scattered fraction from the photoionized material to be $\sim 0.006$. We use this intrinsic power law to estimate the obscuring column density. For densities $>$ 10$^{24}$ cm$^{-2}$, \citet{matt99} show that electron absorption due to Compton scattering can be significant and photoelectric absorption alone is not sufficient. For a covering fraction $\ll$1, electron scattering out of the line-of-sight is the only contributing process, and absorption can be sufficiently modeled by a combination of XSPEC models ``wabs'' and ``cabs''. The unabsorbed power-law parameters were frozen to the values described above while the column density was varied and $\chi^2$ compared with the best-fit value. We found that at column densities less than 3$\times$10$^{24}$ cm$^{-2}$, the $\Delta\chi^2$ due to the presence of the direct component is $>$2.78, so it contributes noticeably to the observed XMM spectrum. Thus the column must be $>3\times$10$^{24}$ cm$^{-2}$. This is so modified by electron scattering that it makes little difference to the extrapolated SWIFT bandpass, so that the total continuum flux (direct + reflected + scattered components; Figure \ref{dummy}) in the 14-195 keV regime is 3$\times$10$^{-12}$ erg cm$^{-2}$ s$^{-1}$. However, the reflector has a large solid angle as inferred from the PEXRAV modeling. In this scenario, electron scattering into the line-of-sight might partially compensate for scattering out of the line-of-sight and the required absorption column density could be higher. For a covering fraction of 1, the direct component may be modeled using PLCABS, developed by \citep{yak97}. For a column density of order 10$^{24}$ cm$^{-2}$, the inferred flux for this model, in the 14-195 keV regime is fairly similar. The difference, and contribution of electron scattering to absorption, becomes more significant for higher column densities (Figure \ref{fig-cvd}). Substituting PLCABS to model the direct component with an absorbing column of 3$\times$10$^{24}$ cm$^{-2}$, gives a somewhat higher flux of 4$\times$10$^{-12}$ erg cm$^{-2}$ s$^{-1}$ in the 14-195 keV regime. The flux values from both models are significantly lower than the SWIFT/BAT sensitivity, consistent with the non-detection of IC 2560.

\section{CONSTRAINING THE SOURCE GEOMETRY}
\subsection{Warm Scatterer}

The detection of line emission at 6.7 keV due to He-like Fe (FeXXV) indicates presence  of a warm, partly-ionized medium. The 6.7 keV feature is intrinsically a composite of 4 emission lines, unresolved in the XMM data: a forbidden line at 6.65 keV, two intercombination lines at 6.67 keV and 6.68 keV and a resonant line at 6.7 keV \citep{mat96}. EPIC does not have sufficient energy resolution to differentiate between these energies. Nevertheless, when allowed to vary, the rest frame energy of the line in the spectral model converged to 6.7$_{-0.2}^{+0.3}$ keV. This may indicate that the resonance feature marginally dominates the total emission of the complex and implies densities less than 10$^{23}$ cm$^{-2}$, in accordance with \citet{mat96}, much less than that for the central source. We conclude the scatterer and central source are well enough separated that we have a relatively clear line-of-sight to the former. We associate this warm gas with the scattered power-law component that dominates the model continuum around 2 keV. In the energy range 2-10 keV, this component contributes 6.7$_{-1.1}^{+1.2}$$\times$10$^{39}$ erg s$^{-1}$ or $\sim$1\% of the the intrinsic emission incident on the scattering medium. We can estimate a lower limit to the column density of the scattering medium from the fraction of scattered flux, 
\begin{equation}
\frac{L_{scattered}}{L_{Intrinsic}} = \frac{\Omega}{4\pi} N_{H} \sigma_{T}
\end{equation}
where $\Omega$/4$\pi$ is the covering fraction, N$_{H}$ is the column density and $\sigma_{T}$ is the Thomson cross-section. We find, ($\Omega$/4$\pi$) N$_{H}$=1.5$\times$10$^{22}$ cm$^{-2}$ for the scattering medium. For a covering fraction $<$ 1, this is a lower limit for the scattering medium.

We can apply the analysis of \citet{bia02} to obtain some order of magnitude estimates of the ionization parameter and distance of the warm scatterer from the central source. The ionization parameter is defined to be \\
\begin{equation}
U_{x} = \frac{\int_{2keV}^{10keV}\frac{L_\nu}{h\nu} d\nu}{4\pi ncr^2}\\
\end{equation} 
where $\it{L}$ is the ionizing luminosity, $\it{n}$ the particle density and $\it{r}$ the distance of the scattering cloud from the source. The equivalent width of the 6.7 keV line, 2.01$^{+1.07}_{-1.07}$ keV measured with respect to the scattered continuum is significantly higher than expected for the above range in estimated column density \citep{bia02}, unless we take into account turbulent motion of the warm gas \citep{bia05a}. The turbulent motion of the scattering material will result in line broadening, giving rise to larger observed equivalent width. Nonetheless, as can be seen from Fig. 4 in \citet{bia02}, the equivalent width is underestimated for all densities, hence adopting the ionization parameter corresponding the the peak in equivalent width for each column density, we estimate U$_{x}$ $\sim$0.32. Though we do not detect the FeXXVI line at 6.955 keV, we obtain an upper limit of 196 eV from the current data. This yields a lower limit on U$_{x}$ of 0.2. For a photon index of 1.75, Equation 2 reduces to $U_{x}=0.23 (L_{2-10}/(nr^2))$. We can express the particle density in terms of the column density of the scatterer and the cloud size $\it{R}$, nr$^2$ $\sim$ $r^2$$(N_H/R)$ from which we obtain $L_{2-10}/U_x = r^2 (N_H/(0.23 R))$. If we assume the cloud size is comparable to its distance from the center, for a covering fraction of 1, $\it{r}$ is on the order of 10pc. For a smaller covering fraction, the column density of the scattering cloud will be higher and distance of the scatterer to the central engine will be smaller. Nonetheless, our characterization of the warm scattering medium is subject to some important caveats. The ionization parameter is particularly uncertain. Specifically, we have assumed the relationship between equivalent width and ionization parameter will be relatively unchanged despite the introduction of a turbulent velocity, and rely on a scaling factor to account for the higher observed equivalent width. 

We note that the observed X-ray spectrum of IC 2560 does not show any intermediate ionization lines, i.e. between the mostly neutral K$\alpha$ lines due to cold Mg, Si, S, Ca, Fe and Ni, and the highly ionized He-like Fe at significantly higher temperatures. Emission lines from diverse ionization states of these and other elements have been detected in other Seyfert 2 sources such as Circinus \citep{mas06} and NGC 1068\citep{pou06}. It is likely that we lack the sensitivity to detect these lines. It is also possible that any intermediate ionization lines are blended with the detected K$\alpha$ lines. This would provide a natural explanation for the relatively high observed equivalent widths. Source geometry might provide a third scenario where the transition region between the warm scatterer and cold reflector is either obscured from our line-of-sight or the scattering and reflecting media are (spatially) well-separated. An optically thick, mostly neutral medium is responsible for the reflection of intrinsic emission, observed above 3 keV while an optically thin, mostly ionized medium, with significantly lower column density scatters intrinsic emission along our line-of-sight. An obscured transition region between the two media, with no detected contribution to the observed spectrum would require a very special line-of-sight. Alternately, the scattering medium may reside above the accretion disk, as would a corona or an outflowing wind. 
\subsection{Reflecting medium and Covering fraction}

The presence of K$\alpha$ emission from mostly neutral species in the reflected emission indicates a low degree of ionization of the reflecting medium. We hypothesize that the cold reflecting medium and the obscuring material along our line-of-sight are one and the same. In that case, the cold reflecting medium has a very high column density and is optically thick. The fluorescence lines detected in the XMM spectrum probably originate in the reflecting medium and aid in better understanding its composition and geometry. The Fe K$\alpha$ line width is consistent with supersolar abundance \citep{ball02} as is the upper limit for the Compton shoulder \citep{mat02}. The observed line widths for the remaining fluorescent lines are also much higher than predicted by \citet{mat97} for a 60$^o$ and 90$^o$ inclination angles, in most cases, by a factor of 2 or more. This may imply a metal-rich environment and/or a different geometry than that discussed by \citet{mat97}. The calculations in \citet{mat97} are most relevant for high angles of incidence and a plane parallel slab illuminated by an external source. A different geometry of source illumination may affect the observed equivalent widths and may explain the high values obtained for lines observed in IC 2560. We cannot rule out chemical enrichment as an alternative. The observed equivalent widths are consistent with 2$\times$Z$_\odot$. However, supersolar abundances of lower metals would result in partial absorption of the higher energy Fe K$\alpha$ photons \citep{geo91}. This implies that the Fe abundance could be $\sim$3-4$\times$Z$_\odot$. Enhancement of low energy K$\alpha$ line widths due to blending with L-shell Fe lines is also likely. In this case, it is possible that the lower elements have solar metallicities and Fe abundance is $\sim$2$\times$Z$_\odot$.

\begin{figure}
\begin{center}
\epsscale{1}[1]
\plotone{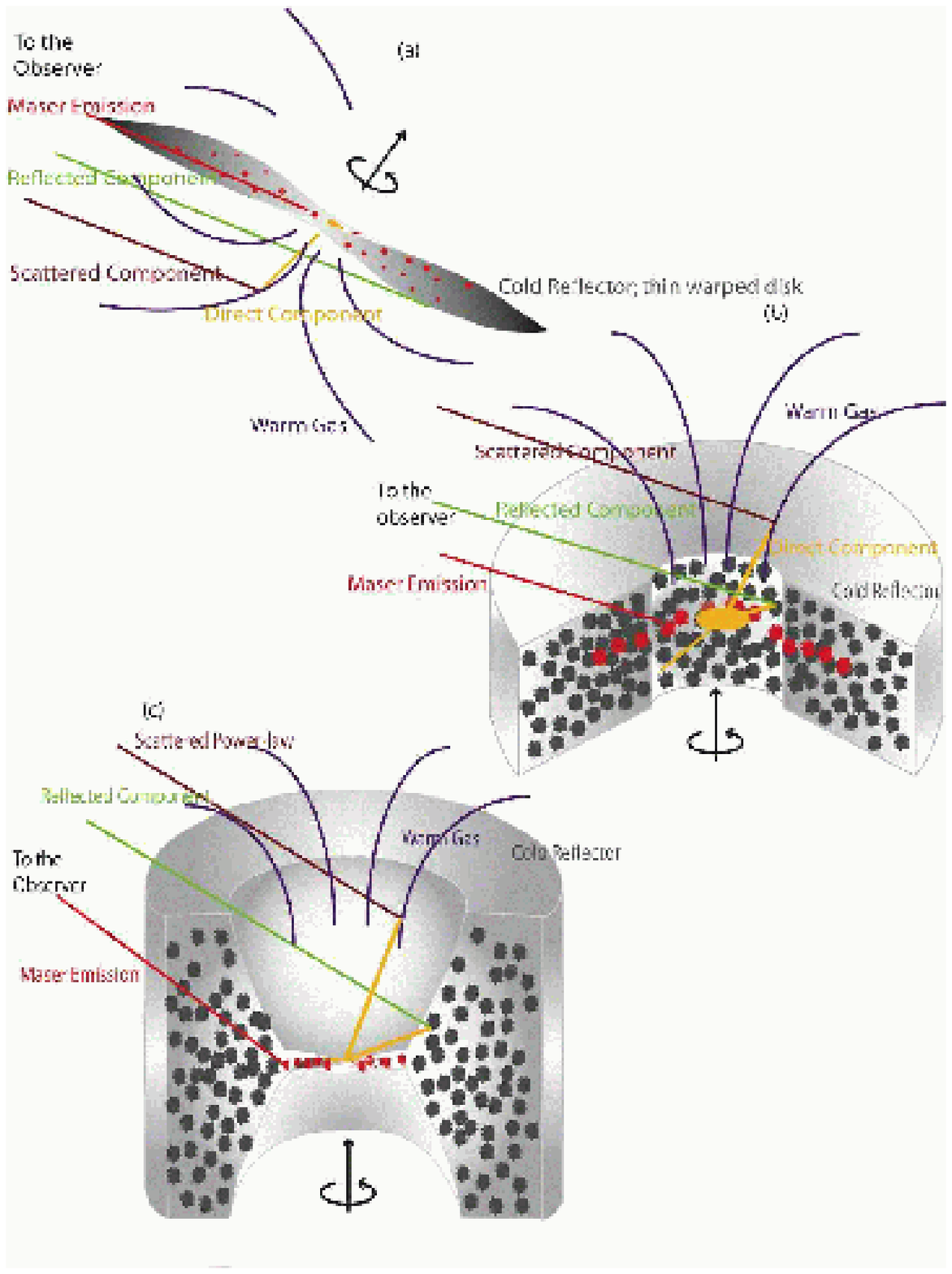}
\epsscale{1}
\caption{Cartoons of possible geometries for AGN in relation to prospective H$_{2}$O maser sources. Figure (a) illustrates a warped, thin disk, (b) illustrates a thick disk, where the masers only trace the disk equatorial plane and (c) shows a flared disk where the disk thickness increases to h$\sim$r beyond the radius of observed maser emission. Cross-sections through the disks and maser distributions are shown for (b) and (c). Though masers may surround the central engines in azimuth, the observer will see only a fraction of the maser emitting clouds depending on line-of-sight and velocity coherence arguments.\label{fig-geom}}
\end{center}
\end{figure}

The 25 \% upper limit on the Compton shoulder obtained from the composite PN and MOS data is consistent with a range of inclination angles, from 30$^{o}$ to 90$^{o}$ \citep{mat02}. We propose three different geometries for the reflector based on current data, where the reflector could be a warped thin disk, a thick clumpy disk or a flared disk inside a clumpy torus (Figure \ref{fig-geom}). The distribution of maser emission is consistent with an edge-on view of a thin Keplerian disk with little or no warp \citep{iwa02}. If the maser emission originates in a relatively flat part of the disk, or our line-of-sight is such that we do not observe the warp, for e.g. orthogonal to a line of nodes, the warped disk scenario may be viable. However, it would be difficult to produce the relatively large (13\%) fraction of the intrinsic emission observed in reflection, from a warped disk. Also the covering fraction for the obscuring medium inferred from modeling (0.75, refer section 4) would imply an extreme warp. Given the special constraints required to reconcile observations, a warped disk seems unlikely. A thick clumpy disk, with a filling factor high enough to obscure the central engine may be a feasible alternative. \citet{nen02} show a clumpy torus structure can produce the necessary obscuration to shield a Seyfert 1 nucleus and produce a Compton thick source. In this case the masers would trace the equatorial plane of the disk toward the inner radius, where pump energy from the central engine can penetrate the clumpy disk as shown by \citet{kar99} in the particular case of NGC 1068. The width of the Fe-line would be dictated by virial rotation. The observed 30 eV intrinsic width is consistent with the estimated central engine dynamical mass of 2.9$\times$10$^{6}$ M$_{\odot}$ and a radius of $\sim$ 0.04 pc, assuming Keplerian rotation. Radiative forces may drive a third scenario with a thin inner disk and inflated outer disk \citep{cha07,kro07} which can also produce the observed reflection fraction, equivalent widths and the intrinsic Fe K$\alpha$ width. The later would be determined by the somewhat slower rotation of the torus and associated turbulent motions in the thickened disk, due to radiative heating, which can be comparable to gravitational forces at these radii.
\subsection{Comparison with other H$_2$O disk maser host galaxies}

We compare IC 2560 to the existing (though small in number) sample of known maser hosts with similar sub-parsec scale edge-on disk structure. The selection criteria were: (1) maser emission originating in a disk as determined from VLBI studies or inferred from maser spectra featuring red and blueshifted, high velocity Doppler components (so that the mass of the black hole can be estimated); (2) high inferred obscuring column density, in excess of $10^{24}$ cm$^{-2}$. We also include NGC 4258 as this is the archetypal maser host galaxy even though it has column of $<10^{23}$ cm$^{-2}$ in contrast to the Compton thick objects in our sample.

Table 4 gives observed details of the eight sources in the sample. IC 2560 is typical of these in many respects such as equivalent width of the Fe K line (NGC 3079 and Circinus), reflected fraction (Mrk 3, Circinus) and Compton shoulder (NGC 1068, Mrk 3 and Circinus). All this suggests a similar geometry for the reflecting and absorbing material in these sources (edge-on geometry where the obscuring/reflecting material is along line of sight; \citet{mat02}). For a zeroth order analysis where we assume thermodynamic equilibrium for these sources, the temperature should be $ 4 \pi R^2 \sigma T^4 \sim L_{bol}$ (both sides also depend in the same way on the covering factor of the material, $\Omega/4\pi$, as this controls both the emission area and the fraction of illumination intercepted). Thus $T^4\propto L_{bol}/R^2\propto L_{2-10}/R^2$ for a constant bolometric correction, which allows us to crudely estimate the temperature for the region of the disk where the maser action occurs. For the observed L$_{2-10 keV}^{I}$/L$_{Edd}$ taken from Table \ref{tab-mas2}, we assume a bolometric correction factor of 20, consistent with the range 15-25 suggested by \citet{vas07}.

The significance of this relation to the data is limited by the large uncertainties in radius and luminosity. Disk geometry (e.g., warping, inclination to the line of sight) and maser beaming can substantively affect the inferred inner radius. We assume an uncertainty of 20\% in the radius due to these factors and use it in weighting the rms scatter. The luminosity estimates depend on complicated model fitting to spectra, which can be particularly difficult for the high obscuration seen in these maser hosts. We cannot quantify the uncertainty in luminosity and instead note that the following analysis should be treated with due caution.

\begin{figure}
\begin{center}
\scalebox{1}[1]{\plotone{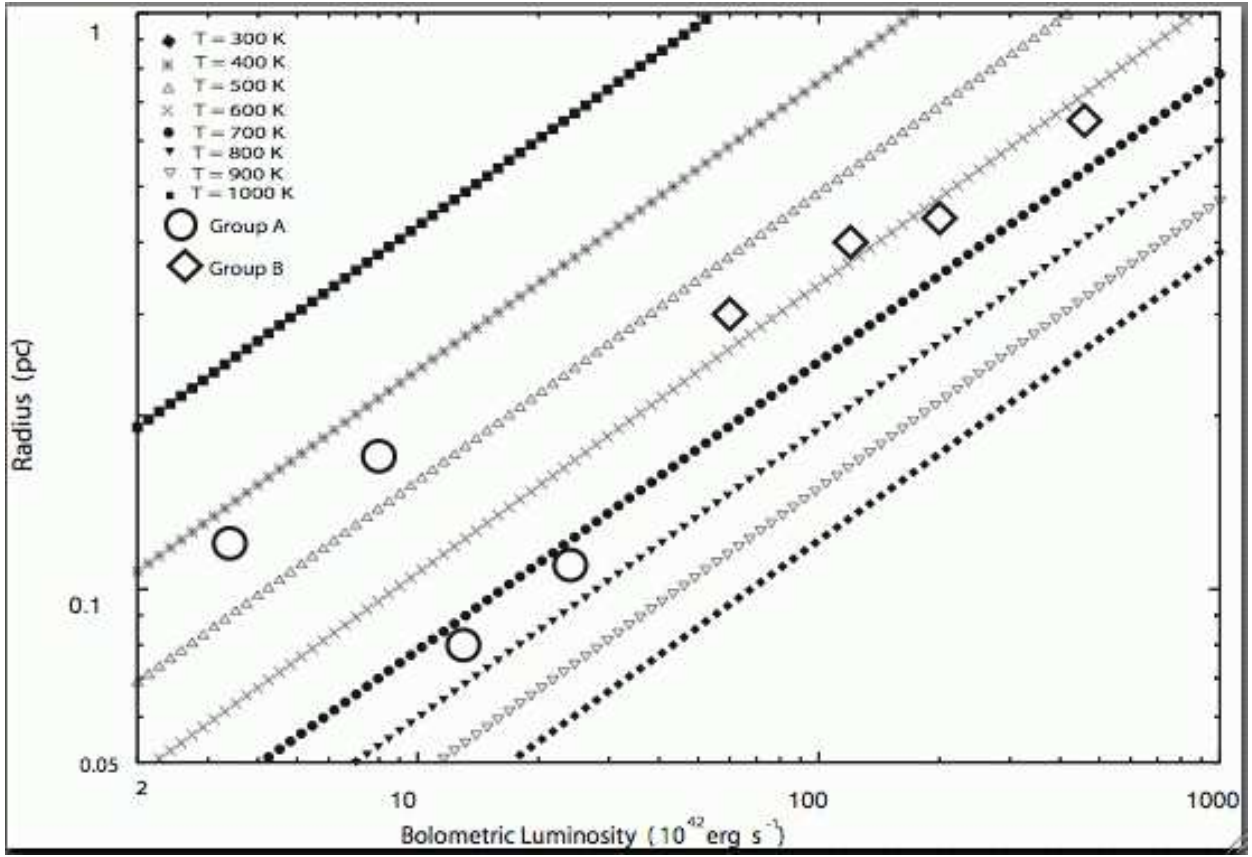}}
\caption{Correlation between observed inner radius of maser emission and modeled intrinsic bolometric luminosity. Data overplotted with the relation anticipated for external radiative disk heating and thermodynamic equilibrium, R $\propto$ L$^{0.5}$ for disk temperatures 300 K, 400 K, 500 K, 600 K, 700 K, 800 K, 900 K and 1000 K. Disks in Group B dominate the higher end of the luminosity distribution.
\label{fig-lir}}
\end{center}
\end{figure}

A limited fit to the data in log-log space can be obtained by including uncertainties in radius alone and yields a curve with best-fit temperature $\sim$600 K.  Figure \ref{fig-lir} shows $R$ plotted against $L_{bol}$, together with the predicted relation for constant temperatures of 300 K, 400 K, 500 K, 600 K, 700 K, 800 K, 900 K  and 1000 K. The observed inner radii of maser emission may be consistent with temperatures in the range 500-700 K, this is in agreement with the predicted temperature range of 400-1000 K for maser disks \citep{eli92}. In physical disks, inefficiency of radiative heating is likely balanced by viscous heating and radiative dissipation, which have not been included in this simplistic analysis \citep{des98}. The temperatures may be even higher for the more luminous sources where the Eddington ratio may be more consistent with a higher bolometric correction of 40-70 \citep{vas07}. Conversely, if the estimated unabsorbed luminosities for the sources in this sample are systematically lower, they will imply smaller disk temperatures. Better understanding of luminosity estimates in this sample and the associated uncertainties are necessary for a more detailed analysis.

Warps are a marked feature of some of the accretion disks traced by water masers, and the degree of warp may also show some correlation with X-ray luminosity.  We divide the sources into two groups, A and B, depending on the distribution of maser emission in the disk. Group A includes IC 2560, Circinus and NGC 3393 all of which show maser emission along a thin disk in the equatorial plane and Keplerian rotation curves, similar to NGC 4258. Group B consists of NGC 1068, NGC 4945, NGC 3079 and NGC 1386 all of which show maser emission from a geometrically thick structure with flattened rotation curves.  We note that the largest warp  in Group A appears in the highest X-ray luminosity galaxy, Circinus \citep{gre03}. IC 2560 and NGC 3393 show X-ray luminosities lower by at least a factor of 2 while we infer  that the warps are smaller by factors of $\la$ 3, from maps of maser distributions in these two sources (Greenhill \etal, 2008, in prep). NGC 4258 exhibits variation in Euler angles that is several times smaller than Circinus for a X-ray luminosity lower by an order of magnitude (Table \ref{tab-mas2}). These observations may be consistent with models in which radiative torques driven by disk irradiation are responsible for warps \citep{pri96,mal96}. On the other hand, the absence of strong warps among AGN in Group B is difficult to understand, we speculate that presence of clumpy disk structure in Group B and the associated large variations in optical to infrared re-radiation may reduce torques that drive the orbits of disk material out of the plane and warp the disk.

We characterize the degree of substructure within the disk using the Toomre Q-parameter \citep{too64}. It is also known as a disk stability criterion and can be used as an indicator of possibility of fragmentation due to self-gravity. For Q$\gg$1, the impact of self-gravity is small and implies stability of a smooth distribution of disk material. For Q$<$1, the reverse is true, and the disk material may be anticipated to clump despite orbital motion in a central potential. Assuming uniform gas disks with observed central mass, at a given distance R,  Q is defined as\\
\begin{equation}
Q = \frac{\Omega v_c}{\pi G \Sigma}
\end{equation}
where $\Omega$ is the angular velocity, v$_c$ is velocity dispersion and $\Sigma$ is the surface mass density. We approximate the dispersion using the mean velocity width spanned by maser clumps along the rotation curves traced in position velocity diagrams. For a disk in pressure equilibrium, the surface density is $<$n$>$m$_{H_2}$ v$_c$/v$_{rot}$ R, where v$_{rot}$ is orbital velocity at the radius R, $<$n$>$ is the number density of the disk gas and m$_{H_2}$ is molecular mass. There is significant uncertainty (up to an order of magnitude) associated with the Q parameter discussed here, mainly due to the range of densities that can support maser action ($10^8$-$10^{10}$ cm$^{-3}$); we adopt 10$^9$ cm$^{-3}$. We find that at the inner radius of maser emission, disks in Group A have a Q$>$1 (Table \ref{tab-mas2}). A similar calculation of the Q-parameter is not appropriate for disks in Group B where maser observations provide significant evidence of self-gravity in the disk (for e.g., NGC 3079: \citet{kon05}, NGC 1068: \citet{lod03}).

If the assumptions we make in computing the Q parameter are broadly valid, the range spanned by Q-parameters for these sources may point to a unified picture of accretion disk structure in maser hosts. Assuming hydrostatic equilibrium, \citet{milo04} suggest a characteristic distance scale of 0.42pc for a central mass of the order of 10$^7$ $M_\odot$ beyond which the disk may be marginally unstable and likely to form clumpy structures due to self-gravity. In the unified picture, the overall disk structure in all these sources could be the same, with the maser emission in Group A sources tracing the inner, gravitationally stable parts of accretion disks and gravitationally unstable outer parts in Group B (Table \ref{tab-mas2}, Figure \ref{fig-lir}). We focus on the inner radii of the maser disks, which we may anticipate to be linked to intrinsic luminosities via the temperature of dust sublimation,$\sim$10$^3$K (\cite{neu94}; the outer radii of the maser disks, may be determined by the contribution of other factors such as density profiles, shielding and re-emission from surrounding gas). The lower luminosity of sources in Group A allows survival of molecules (and maser action) at smaller radii, where the disk structure is well-ordered. Among higher luminosity sources in Group B, molecular emission arises farther from the black hole, which is also where the disk structure is likely to be clumpy. X-ray observations of other Compton thick  maser hosts, where VLBI data points to a disk origin will help expand the sample size and test the preliminary trend reported here.
\section{SUMMARY}

A deep 70 ks XMM-Newton spectrum for IC 2560 shows evidence for a warm scattered continuum component as well as number of emission lines not detected in earlier Chandra observations. These include K$\alpha$ emission from Mg (1.25 keV), Ca (3.6 keV) and Ni (7.47 keV) as well as Fe K$\beta$ (7.05 keV) and K$\alpha$ line emission from Fe XXV (6.7 keV). We do not detect line emission from species at intermediate ionization states, between the nearly neutral material that gives rise to the K$\alpha$ emission, and the warm scattering medium with FeXXV. The direct X-ray emission from the central source appears to be completely obscured up to 10 keV; the inferred column density is in excess of a few times 10$^{24}$ cm$^{-2}$. This is in agreement with a 25\% upper limit on the magnitude of the Compton shoulder. The reflecting medium is mostly neutral, optically thick and could also provide the obscuring column toward the central engine.  The 6.7 keV line is associated with the optically thin, mostly ionized, scattering medium,  with significantly lower density. The dominance of the resonance line at 6.7 keV indicates the density has to be lower than 10$^{23}$ cm$^{-2}$, while the estimated fraction of intrinsic emission scattered along line-of-sight indicates a lower limit of 10$^{22}$ cm$^{-2}$. The scattering medium can plausibly exist on the scale of a few parsec, but the exact geometry is unknown. Limits on density of the scatterer indicate a covering fraction between 0.1 and 1. Using a small sample of H$_{2}$O disk maser hosts, selected using physically motivated criteria, we find that the unobscured luminosity of the central source may shape molecular disk structure in this sample. The inner radius as well as degree of substructure in the disk may be related to the unobscured X-ray emission, though a larger sample is required to confirm this trend.

\acknowledgements
We thank E.M.L. Humphreys for many useful discussions and Julian Krolik for helpful comments. This research has made extensive use of the NASA/IPAC Extragalactic Database(NED) which is operated by the Jet Propulsion  Laboratory (JPL), California Institute of Technology, under contract with NASA, and the NASA's Astrophysics Data System Bibliographic Services. This work was supported in part by NASA grant NNG05GK24G, Department of Energy contract to SLAC no.DE-AC3-76SF00515, and by NASA observing grant NNX06AE19G. CD acknowledges support from a PPARC Senior fellowship.

\clearpage

\begin{deluxetable}{llccccccccc}
\rotate
\tabletypesize{\scriptsize}
\tablecolumns{10}
\tablewidth{0pt}
\tablecaption{Characteristics of XMM EPIC PN+MOS spectrum}
\tablehead{\colhead{Model name}&\colhead{Model} &\colhead{N$_{H}$\tablenotemark{(a)}}& \colhead{kT[1]} & \colhead{kT[2]} & \colhead{kT[3]}&\colhead{kT[4]}& \colhead{$\Gamma$} & \colhead{Eq. W.\tablenotemark{(b)}} &\colhead{$\chi^2$/dof}\\
& & (10$^{20}$cm$^{-2}$)& (keV)& (keV)& (keV)& (keV)& & (keV)& \\}
\startdata\\
Chandra-1& ab(pexrav+2mekal+3gauss)\tablenotemark{(c)}& 1.8$_{-0.2}^{+0.3}$& 0.63$_{-0.02}^{+0.01}$& 0.08$_{-0.00}^{+0.01}$\tablenotemark{(d)}& ...& ...&2.18$_{-0.13}^{+0.01}$& 2.40$_{-0.17}^{+0.17}$& 493.2/317\\
\\
\tableline
\\
XMM-1& ab(pexrav+3mekal+8gauss)\tablenotemark{(e)}& 1.6$_{-0.2}^{+0.2}$& 0.77$_{-0.05}^{+0.05}$& 0.08$^{+0.01}_{-0.00}$\tablenotemark{(d)}& 0.32$^{+0.03}_{-0.03}$& ...& 2.29$_{-0.13}^{+0.13}$& 2.70$^{+0.18}_{-0.18}$& 358.7/312\\
\\
\tableline
\\
XMM-2& ab(pexrav+4mekal+8gauss)\tablenotemark{(f)}& 14.7$^{+0.2}_{-0.2}$& 0.71$^{+0.06}_{-0.07}$& 0.08$^{+0.01}_{-0.00}$\tablenotemark{(d)}& 0.31$_{-0.05}^{+0.04}$& 4.4$_{-1.7}^{+2.9}$& 1.62$^{+0.21}_{-0.20}$& 2.70$^{+0.18}_{-0.18}$&322.0/311\\
\\
\tableline
\\
XMM-bf& ab(pexrav+powerlaw+3mekal+8gauss)\tablenotemark{(g)}& 14.7$_{-2.1}^{+2.3}$& 0.73$_{-0.09}^{+0.06}$& 0.08$_{-0.00}^{+0.01}$\tablenotemark{(d)}& 0.31$_{-0.06}^{+0.03}$& ...& 1.75$_{-0.23}^{+0.21}$& 2.32$_{-0.17}^{+0.18}$& 317.4/311\\
\\
\enddata
\tablenotetext{(a)}{The lower limit on the absorption column density fixed to Galactic absorption.}
\tablenotetext{(b)}{Equivalent Width for Fe K$\alpha$ line, with respect to the reflected continuum.}
\tablenotetext{(c)}{Includes contributions from line emission corresponding to Si, S, and Fe  K$\alpha$ at energies 1.75 keV, 2.31 keV and 6.4 keV as well as continuum emission from the reflected component and 2 MeKaL plasmas.}
\tablenotetext{(d)}{Parameter value converged to lower limit.}
\tablenotetext{(e)}{Includes contributions from line emission corresponding to Mg, Si, S, Ca, Fe K$\alpha$, Fe K$\beta$ and He-like Fe K$\alpha$ at energies 1.25 keV, 1.75 keV, 2.31 keV, 3.6 keV, 6.40 keV, 7.05 keV and 6.7 keV respectively as well as continuum emission from the reflected component and 3 MeKaL plasmas.}
\tablenotetext{(f)}{Includes contributions from line emission corresponding to Mg, Si, S, Ca, Fe and Ni  K$\alpha$ at energies 1.25 keV, 1.75 keV, 2.31 keV, 3.6 keV, 6.4 keV and 7.47 keV; Fe K $\beta$ at 7.05 keV as well as continuum emission from the reflected component and 4 MeKaL plasmas.}
\tablenotetext{(g)}{Includes contributions from line emission corresponding to Mg, Si, S, Fe and Ni K$\alpha$ at energies 1.25 keV, 1.75 keV, 2.31 keV, 6.4 keV, 7.47 keV; Fe K $\beta$ at 7.05 keV and He-like Fe K$\alpha$ at 6.7 keV as well as continuum emission from the reflected component, 3 MeKaL plasmas and a scattered powerlaw, with photon index same as the reflection model.}  
\label{tab-spec}
\end{deluxetable}

\clearpage

\begin{deluxetable}{lcccccccc}
\rotate
\tabletypesize{\scriptsize}
\tablecolumns{11}
\tablewidth{0pt}
\tablecaption{Line Equivalent Widths for XMM and Chandra data}
\tablehead{\colhead{Instrument} & \colhead{Mg K $\alpha$}& \colhead{Si K $\alpha$}& \colhead{S K $\alpha$}& \colhead{Ca K $\alpha$}& \colhead{Fe K $\alpha$}& \colhead{Fe K $\beta$}& \colhead{Ni K $\alpha$} & \colhead{He-like Fe K $\alpha$}\\
&1.25 keV\tablenotemark{(a)}& 1.75 keV\tablenotemark{(a)}& 2.31 keV\tablenotemark{(a)}& 3.6 keV\tablenotemark{(a)}& 6.4 keV\tablenotemark{(a)}& 7.05 keV\tablenotemark{(a)}& 7.47 keV\tablenotemark{(a)}& 6.7 keV\tablenotemark{(a)}\\
&(keV)& (keV)& (keV)& (keV)& (keV)& (keV)& (keV)& (keV)\\
}
\startdata
\\
XMM PN+MOS\tablenotemark{(b)} & 0.55$_{-0.24}^{+0.23}$& 0.55$_{-0.15}^{+0.14}$& 0.34$_{-0.16}^{+0.15}$& 0.08$_{-0.08}^{+0.08}$& 2.32$_{-0.16}^{+0.18}$& 0.30$_{-0.02}^{+0.02}$& 0.31$_{-0.02}^{+0.02}$& 2.01$_{-1.07}^{+1.07}$\tablenotemark{(c)}\\
\\
Chandra ACIS-S\tablenotemark{(d)}& -& 0.33$_{-0.17}^{+0.17}$& 0.34$_{-0.20}^{+0.19}$& ...& 2.77$_{-0.49}^{+0.49}$& ...& ...& ...\\
\\
\enddata
\tablenotetext{(a)}{Rest frame energies.}
\tablenotetext{(b)}{Equivalent widths for XMM data using best-fit model XMM-bf, described in Table 1, calculated with respect to reflected component.}
\tablenotetext{(c)}{Equivalent width calculated with respect to scattered component of the continuum.}
\tablenotetext{(d)}{Equivalent widths for Chandra data using best-fit model from \citet{mad06}, calculated with respect to reflected component.}
\label{tab-eqw}
\end{deluxetable}

\begin{deluxetable}{lccc}
\tabletypesize{\scriptsize}
\tablecolumns{6}
\tablewidth{0pt}
\tablecaption{Emission Characteristics from Chandra and XMM-Newton Studies\tablenotemark{(a)}}
\tablehead{
\colhead{Energy range} &\colhead{Emission} &\colhead{Chandra} & \colhead{XMM-Newton}\\}
\startdata
6.4 keV& Equivalent Width& 2.77$_{-0.49}^{+0.49}$ keV & 2.32$_{-0.16}^{+0.18}$ keV\\
\\
6.25-6.45 keV& Observed Flux& $1.35_{-0.50}^{+0.45}$$\times$10$^{-13}$ erg cm$^{-2}$ s$^{-1}$& $1.23_{-0.09}^{+0.09}$$\times$10$^{-13}$ erg cm$^{-2}$ s$^{-1}$\\
\\
& Observed Luminosity& $2.85^{+0.95}_{-1.18}$$\times$10$^{40}$ erg s$^{-1}$& $2.68_{-0.29}^{+0.09}$$\times$10$^{40}$ erg s$^{-1}$\\
\\
2-10 keV& Observed Flux& 3.84$_{-0.46}^{+2.11}$$\times$10$^{-13}$ erg cm$^{-2}$ s$^{-1}$& 3.88$_{-0.51}^{+0.18}$$\times$10$^{-13}$ erg cm$^{-2}$ s$^{-1}$\\
\\
&Observed Luminosity& $8.06_{-3.05}^{+1.32}$$\times$10$^{40}$ erg s$^{-1}$& $8.15_{-0.94}^{+0.31}$$\times$10$^{40}$ erg s$^{-1}$\\
\enddata
\tablenotetext{(a)}{Values and ranges from model Chandra-1, fitted to Chandra data, described by \citet{mad06}, and XMM PN+MOS data as described in Table 1.}
\label{tab-var}
\end{deluxetable}

\clearpage

\begin{deluxetable}{ccccccccc}
\rotate
\tabletypesize{\scriptsize}
\tablecolumns{9}
\tablewidth{0pt}
\tablecaption{Observed Properties of Compton-thick Disk Maser Nuclei}
\tablehead{
& (a)& (b)& (c)& (d)& (e)& (f)& (g) &\\
\colhead{Source}& \colhead{M$_{SMBH}$}& \colhead{Disk size} & \colhead{D$_{L}$}&\colhead{N$_H$}& \colhead{L$_{Obs}$(2-10)}& \colhead{Eq. W.}& \colhead{CS}&\colhead{References}\\
& (M$_\odot$)& (pc) & (Mpc)& (cm$^{-2}$)& (erg s$^{-1}$)& (keV)& (\%) &\\}
\startdata
IC 2560& 2.9$\pm$0.6$\times$10$^{6}$& 0.08-0.27&  41.4&$>$ 3$\times$10$^{24}$& 8.2$\times$10$^{40}$ & 2.31$_{-0.16}^{+0.18}$& $<$25\tablenotemark{(h)}& (a),(b)- 1.\\
& & & & & & & & (d),(e),(f), (g)-this paper\\
\tableline
& & & & & & & & (a),(b)-2.\\
NGC 4945& $\sim$1.4$\times$10$^{6}$ & $\sim$0.3& 3.7& $\sim$4$\times$10$^{24}$& 3$\times$10$^{38}$& 1.6$_{-0.1}^{+0.1}$& ...&  (d)-3.\\
& & & & & & & & (e)-4.\\
& & & & & & & & (f)-5.\\
\tableline
& & & & & & & & (a),(b)-6.\\
NGC 1068& $\sim$2$\times$10$^{7}$& 0.65-1.1 & 14.4& $\ge$10$^{26}$& 5.2$\times$10$^{40}$& 1.2& $\sim$20&  (d)-7.\\
& & & & & & & & (e),(g)-8.\\
& & & & & & & & (f)-9.\\
\tableline
& & & & & & & &  (a)-10.\\
Mrk 3&$\sim$ 4.5$\times$10$^{8}$& ...& 57.7& 1.36$_{-0.04}^{+0.03}$$\times$10$^{24}$& 1.4$\times$10$^{42}$\tablenotemark{(i)}& 0.61$_{-0.05}^{+0.03}$& 10$^{+9}_{-6}$& (d)-11.\\
& & & & & & & & (e),(f),(g)-12.\\
\tableline
& & & & & & & & (a),(b)-13.\\
Circinus& 1.7$\pm$0.3$\times$10$^{6}$& 0.11-0.4& 4.1& 4.3$_{-0.7}^{+0.4}$$\times$10$^{24}$& $\sim$10$^{40}$\tablenotemark{(j)}& 2.25$_{-0.30}^{+0.26}$& 20$\pm$3& (d)-14.\\
& & & & & & & & (e)-15.\\
& & & & & & & & (f)-16.\\
& & & & & & & & (g)-17.\\
\tableline
NGC 3393& 3.1$\pm$0.2$\times$10$^{7}$& 0.17$\pm$0.02 & 53.3& $\sim$4$\times$10$^{24}$& 3.1$_{-0.1}^{+0.2}$$\times$10$^{40}$& 1.4$\pm$0.8& ...& (a),(b)-18\\
& & & & & & & & (d),(e),(f)-19.\\
\tableline
NGC 3079& 2$\times$10$^{6}$& $\sim$0.4& 17.3& $\sim$10$^{25}$& $\sim$10$^{40}$& 2.4$_{-1.5}^{+2.9}$& ...& (a),(b)-20.\\
& & & & & & & & (d),(e),(f)-21.\\
\tableline
& & & & & & & &(a),(b)-22\\
NGC 1386& 1.2$\times$10$^{6}$& 0.44-0.94& 17\tablenotemark{(k)}& $>$2.2$\times$10$^{24}$& 4.9$\pm$0.9$\times$10$^{39}$& 1.8$_{-0.3}^{+0.4}$& ...& (d),(e),(f)-19.\\
\\
\tableline
\tableline
& & & & & & & & (a),(b)-23.\\
NGC 4258\tablenotemark{(l)}& 3.9$\pm$0.3$\times$10$^{7}$& 0.12-0.28\tablenotemark{(m)}& 7.2&6.0-15 $\times$10$^{22}$& 2.4-10.4$\times$10$^{39}$& $\le$0.049& & (d),(e),(f)-24.\\
& & & & & & & & \\
\enddata
\tablenotetext{(c)}{\ Luminosity distance used in calculation of mass, radius and luminosity for each source}
\tablenotetext{(a),(b),(e)}{Uncertainties are quoted where they are known. With the exception of NGC 4258, the mass, radius and luminosity in each object are uncertain by uncertainties in distance, as well as any modeling uncertainties.}
\tablenotetext{(e)}{Observed luminosity in 2-10 keV energy range.} 
\tablenotetext{(f)}{Eq. W. of Fe K$\alpha$ line measured with respect to the reflected component of the spectrum.} 
\tablenotetext{(g)}{Compton shoulder strength relative to the Fe K$\alpha$ line.}
\tablenotetext{(h)}{Upper limit obtained by considering only Pattern 0 events.} 
\tablenotetext{(i)}{Calculated from quoted flux in \citet{bia05b}.}
\tablenotetext{(j)}{Wilson, private communication.}
\tablenotetext{(k)}{Distance adopted from \citet{hen05}}
\tablenotetext{(l)}{NGC 4258 is a Compton-thin source and has been included here only for comparison.}
\tablenotetext{(m)}{Source luminosity range observed over 10 years, variation due to variability in absorption column density.}
\tablerefs{ (1) Greenhill \etal 2007 (in preparation), (2)-\citet{gree97}, (3)-\citet{don03}, (4)-\citet{gua00},(5)-\citet{schu02}, (6)-\citet{gre97}, (7)-\citet{matt97}, (8)-\citet{pou06}, (9)-\citet{mat04}, (10)-\citet{woo02}, (11)-\citet{pou05}, (12)-\citet{bia05b}, (13)-\citet{gre03}, (14)-\citet{mat99}, (15)-\citet{smi01}, (16)-\citet{mat00}, (17)-\citet{mol03}, (18)-\citet{kon06b}, (19)-\citet{gua05},(20)-\citet{kon05}, (21)-\citet{iyo01}, (22)-\citet{bra96}, (23)-\citet{arg07} and references therein, (24)-\citet{fru05}. }
\label{tab-mas1}
\end{deluxetable}

\begin{deluxetable}{cccccccc}
\tabletypesize{\scriptsize}
\tablecolumns{6}
\tablewidth{0pt}
\tablecaption{Inferred Properties of Compton-thick Disk Maser Nuclei}
\tablehead{
\colhead{Source} &\colhead{L$_{2-10}^{I}$\tablenotemark{(a)}} &\colhead{L$_{2-10}^{I}$/L$_{Edd}$} & \colhead{R/R$_{G}$\tablenotemark{(b)}}& \colhead{Q$_{in}$\tablenotemark{(c)}}& \colhead{Q$_{out}$\tablenotemark{(d)}}& \colhead{Group\tablenotemark{(e)}}\\
& (erg s$^{-1}$)& & & & &\\}
\startdata
NGC 4258& 4.2-17.4$\times$10$^{40}$\tablenotemark{(f)}& 1-4$\times$10$^{-5}$& 3.2-7.2$\times$10$^4$& 144&11& A\\
\\
NGC 3393& 4$\times$10$^{41}$\tablenotemark{(g)}& 1$\times$10$^{-4}$& 5.7$\pm$0.4$\times$10$^4$& - &32& A\\
\\
IC 2560& 6.3$\times$10$^{41}$\tablenotemark{(h)}& 2$\times$10$^{-3}$& 2.8-9.5$\times$10$^5$& 36& 0.9& A\\
\\
Circinus& 1.2$\times$10$^{42}$\tablenotemark{(i)}& 5$\times$10$^{-3}$& 6.7-25$\times$10$^5$& 8& 0.2& A\\
\\
NGC 1068& 2.3$\times$10$^{43}$\tablenotemark{(j)}& 1$\times$10$^{-2}$& 3.4-5.8$\times$10$^5$& -& -& B\\
\\
NGC 4945& 3$\times$10$^{42}$\tablenotemark{(k)}& 2$\times$10$^{-2}$& 2.2$\times$10$^6$& -& -& B\\
\\
NGC 3079& 6-12$\times$10$^{42}$\tablenotemark{(l)}& 2-4$\times$10$^{-2}$& 2.1$\times$10$^6$& -& -& B\\ 
\\
NGC 1386& 9.6-38.2$\times$10$^{42}$\tablenotemark{(m)}& 6-24$\times$10$^{-2}$& 3.7-7.9$\times$10$^{6}$& -& -& B\\
\\
\enddata
\tablenotetext{(a)}{Intrinsic (unabsorbed) luminosity in the 2-10 keV range, after accounting for local and Galactic absorption.}
\tablenotetext{(b)}{Radius for maser emission, in gravitational units.}
\tablenotetext{(c),(d)}{Q parameter computed for the inner and outer radii of maser emission for uniform gas disks assuming, particle density of 10$^{9}$ cm$^{-3}$ characteristic of maser action from H$_{2}$O at 22 GHz, and velocity dispersion, v$_{c}$=1.5$\times$10$^{3}$ km s$^{-1}$ corresponding to the sound speed of molecular gas.}
\tablenotetext{(e)}{Grouping based on disk morphology, Group A: Thin, fairly flat disks, Group B: Thick disks.}
\tablenotetext{(g)}{\ Luminosity from reflection dominated modeling of Beppo-Sax data by \citet{sal97}, preferred over transmission dominated modeling, based on analysis of PN data by \citet{mai98}.}
\tablenotetext{(l)}{Intrinsic luminosity quoted by \citep{kon06a}, using PIMMS, does not take into account obscuration due to Compton scattering and may be significantly underestimated.}
\tablerefs{(f)-\citet{fru05}, (g)-\citet{sal97}, (h)-this paper, (i)-\citet{smi01}, (j)-\citet{pou06}, (k)-\citet{gua00}, (l)-\citet{iyo01}, (m)-\citet{kon06a} }
\label{tab-mas2}
\end{deluxetable}

\clearpage

\clearpage
\clearpage
\clearpage

\end{document}